\documentclass[onecolumn]{mn2e}
\usepackage{times}
\input{psfig.sty}

\newif\ifAMStwofonts

\begin{document}

\title[Identifying young GRB fossils]{Identifying young gamma-ray
burst fossils} \author[Ramirez-Ruiz]{Enrico Ramirez-Ruiz$^{1,2}$ \\
$^{1}$School of Natural Sciences, Institute for Advanced Study,
Einstein Drive, Princeton, NJ 08540, USA.\\ $^{2}$Chandra Fellow.}

\date{}

\maketitle

\label{firstpage}

\begin{abstract}
The recent reports of temporal and spectral peculiarities in the early
stages of some afterglows suggest that we may be wrong in postulating
a central engine which becomes dormant after the burst itself. A
continually decreasing postburst relativistic outflow, such as put out
by a decaying magnetar, may continue to be emitted for periods of days
or longer, and we argue that it can be efficiently reprocessed by the
ambient soft photon field radiation. Photons produced either by the
postexplosion expansion of the progenitor stellar envelope or by a
binary companion provide ample targets for the relativistic outflow to
interact and produce high energy $\gamma$-rays. The resultant signal
may yield luminosities high enough to be detected with the recently
launched {\it Integral} and the {\it Glast} experiment now under
construction. Its detection will surely offer important clues for
identifying the nature of the progenitor and possibly constraining
whether some route other than single star evolution is involved in
producing a rapidly rotating helium core which in turn, at collapse,
triggers a burst.
\end{abstract}


\section{Introduction}
The typical GRB model assumes that the energy input episode is brief,
typically $t_{\rm grb} \le 1- 10^2$s (see M\'esz\'aros 2002 for a
recent review). However, peculiarities in the early stages of some
afterglows, e.g. GRB 021004, have served as motivation for considering
a more extended input period in which the energy injection continues
well beyond $t_{\rm grb}$ (Fox et al. 2003). The enduring activity
could in principle emanate from the sluggish drain of orbiting matter
into a newly formed black hole (e.g. MacFadyen \& Woosley 1999) or
from a spin down millisecond super-pulsar (e.g. Usov 1994), which
could produce a luminosity that was, still one day after the burst, as
high as $L \sim 10^{47}$ erg s$^{-1}$.  Collapsar (Woosley 1993,
Paczy\'nski 1998; MacFadyen \& Woosley 1999; Aloy et al. 2000) or
magnetar-like GRB models (Usov 1994; Thompson 1994, Wheeler et
al. 2000) provide a natural scenario for a sudden burst succeeded by a
more slowly decaying energy release (Rees \& M\'esz\'aros 2000;
Ramirez-Ruiz, Celotti \& Rees 2002).

The power output would be primarily in a magnetically driven
relativistic wind, which would be hugely super-Eddington during the
time scales discussed here. Its luminosity may not dominate the
afterglow continuum (Dai \& Lu 1998; Zhang \& M\'esz\'aros 2001), but
we argue that it could be efficiently reprocessed into $\gamma$-rays.
This could be due to the interaction of the postburst relativistic
outflow with the dense soft photon bath arising either from a stellar
companion or from the postexplosion expansion of the remnant shell or
supernova (Hjorth et al. 2003; Stanek et al. 2003). The detection of
such scattered hard $\gamma$-ray radiation would offer the possibility
of diagnosing the nature of the precursor star and the compact object
that triggers the burst.

\section{binary system with a decaying magnetar}

In the generic pulsar model the field is assumed to maintain a steady
value, and the luminosity declines as the spin rate slows
down. However, during the early stages, the magnetic field strength
might decline more rapidly than the slowing down timescale. The power
output in this case declines in proportion to $B^2$. The spindown law
is given by $-I\Omega\dot\Omega= {B^2R^6\Omega^4}/{(6c^3)} +
{32GI^2\varepsilon^2\Omega^6}/{(5c^5)}$ (Shapiro \& Teukolsky 1983),
where $\Omega$ is the angular frequency, $B$ is the dipolar field
strength at the poles, $R$ is the radius of the light cylinder, $I$ is
moment of inertia, and $\varepsilon$ is ellipticity of the neutron
star. The above decay solution includes both electromagnetic (EM) and
gravitational wave (GW) losses. At various times the spin-down will be
dominated by one of the loss terms, and one can get approximate
solutions. When EM dipolar radiation losses dominate the spin-down, we
have $\Omega=\Omega_0(1+t/t_{\rm m})^{-1/2}$. Here
\begin{equation}
t_{\rm m}={3c^3I \over B^2R^6\Omega_0^2}\simeq 10^3~{\rm s}~
I_{45}B_{15}^{-2}P_{0,-3}^2R_6^{-6} 
\label{tm}
\end{equation}
is the characteristic time scale for dipolar spindown (e.g Dai \& Lu
1998), $B_{15}=B/(10^{15}{\rm G})$, $P_0$ is the initial rotation
period in milliseconds. When GW radiation losses dominate the
spin-down, the evolution is $\Omega \approx \Omega_0(1+t/t_{\rm
  gw})^{-1/4}$, where $t_{\rm gw} \simeq 1 {\rm s}
I_{45}^{-1}P_{0,-3}^{4}(\varepsilon/0.1)^{-2}$. GW spindown is
important only when the neutron star is born with an initial $\Omega_0
\ge \Omega_{\ast} \sim 10^4 {\rm s}^{-1}$ (e.g. Blackman \& Yi 1998).
When $\Omega_0 < \Omega_{\ast}$, the continuous injection luminosity
is given by
\begin{equation}
L_{\rm m}(t) =L_{{\rm m},0}(1+ t/t_{\rm m})^{-2}, 
\label{spin}
\end{equation}
where 
\begin{equation}
L_{{\rm m},0}={I\Omega_0^2 \over 2t_{\rm m}} \simeq 10^{49} {\rm
erg~s^{-1}}B_{15}^2P_{0,-3}^{-4}R_6^6.
\label{lm}
\end{equation}
If the neutron star is born with $\Omega_0 > \Omega_{\ast}$, the
timescale for GW-dominated regime is short so that $\Omega$ will be
damped to below $\Omega_{\ast}$ promptly in a time $t_\ast
=[(\Omega_0/\Omega_{\ast})^{4}-1]t_{\rm gw}$. After $\Omega <
\Omega_{\ast}$, GW losses decrease sharply, and the spin-down becomes
dominated by the EM losses. The injection luminosity can therefore be
divided into two phases, i.e. $L=L_{{\rm m},0}/(1+t/t_{\rm gw})$ for
$t < t_\ast$ or $L=L_{{\rm m},\ast}/[1+(t-t_\ast)/t_{m,\ast}]^2$
otherwise, where $L_{{\rm m},\ast} = I\Omega_\ast^2/(2t_{\rm m})$ and
$t_{m,\ast}=3c^3I/(B^2 R^6 \Omega_\ast)$.

The bulk of the magnetar energy $L_{\rm m}$ would be primarily in the
form of a magnetically driven, highly-relativistic wind consisting of
$e^{-}$, $e^+$ and probably heavy ions with $L_\Omega \simeq \zeta
L_{\rm m}$ and $\zeta \le 1$. We envisage that the burst is triggered
by the collapse of a massive star whose helium core is presumed to be
kept rapidly rotating by spin-orbit tidal interactions with its binary
companion.  The relativistic (postburst) wind would then escape the
compact remnant while interacting with the soft photon field of the
companion with typical energy $\theta=kT/(m_ec^2) \sim
10^{-6}-10^{-5}$ (i.e. optical and UV frequencies). The scattered
photons whose energy is boosted by the square of the bulk Lorentz
factor of the magnetized wind (i.e. $\Theta=2\gamma^2 \theta$)
propagate in a narrow $\gamma^{-1}$ beam owing to relativistic
aberration. In the case of a mono-energetic isotropic magnetar wind
(which is likely to be undisturbed even in the presence of a strong
mass outflow from the binary companion; i.e.  $L_\Omega/c \gg
v\dot{M}$), and assuming that the relativistic particles and soft
photons are emitted radially from their source, the energy extracted
from a relativistic particle by the inverse Compton process in the
Thompson regime is given by $d\gamma (r,\varphi) \simeq - L_\ast
\sigma_T \gamma^2 (1-\cos \omega)^2 dr /(4\pi m_e \delta^2 c^3)$,
where $L_\ast$ is the luminosity of the stellar companion, $\omega$ is
the angle between the photon and particle directions before
scattering, and $r$ denotes the distance to the magnetar from the
volume region where the $\gamma$-ray radiation is generated. Here
$\delta=\sqrt{\Lambda^2 \sin^2 \varphi + (r-\Lambda \cos\varphi)^2}$
defines the distance from this volume to the companion, $\Lambda$ the
binary separation, $\varphi$ the angle between the lines connecting
the binary members and the observer, and $\delta \cos \omega=r-\Lambda
\cos\varphi$.
\begin{figure*}
\centerline{\psfig{figure=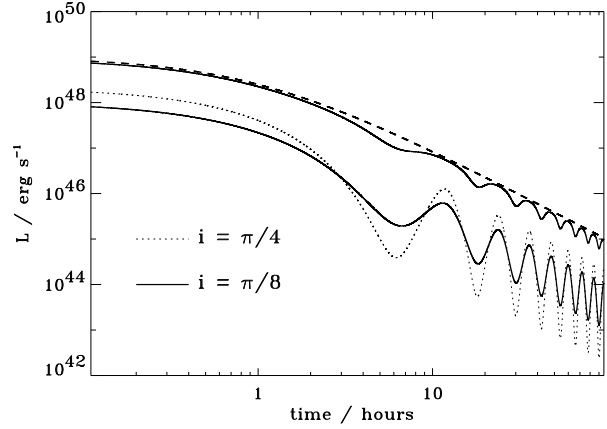,angle=0,width=0.55\textwidth}}
\vspace{-0.5cm} {\caption{The luminosity emitted by the (postburst)
relativistic outflowing wind through the Compton-drag process. The
power output of the magnetically driven relativistic wind declines as
the spin rate slows down (see eq. \ref{spin}) and $B$ is assumed to
maintain a steady value ({\it dashed line}). $L_{{\rm m},0} \simeq
10^{49} {\rm erg~s^{-1}}$, $t_{\rm m} \simeq 10^3~{\rm s}$, and
$\zeta=1$. The luminosity of the $\gamma$-ray emission generated as
the relativistic outflow interacts with the soft photon field of the
companion is shown for $\varsigma$=1,10 and for two different values
of the inclination angle (i.e. $i$). The period of the binary in the
circular orbit is $\tau_p =$ 12 hours.  The emitted luminosity was
derived under the assumption that the relativistic wind is both steady
and mono-energetic. In reality, the properties of the continuing power
output could be more complicated. Such effects will certainly enrich
the temporal and spectral dependence of the resultant $\gamma$-ray
emission.}
\label{fig1}}
\end{figure*} 
The decrease of the particle's Lorentz factor from the magnetar to
infinity is
\begin{equation}
{\Delta \gamma \over \gamma} \simeq 1-(1+ \varsigma \Psi)^{-1},
\label{deltag}
\end{equation}
where
\begin{equation}
\varsigma={\gamma \sigma_T L_\ast \over 4\pi \Lambda m_ec^3} \sim 30
\left({\gamma \over 10^6}\right) \left({L_\ast \over 10^3
L_\odot}\right) \left({\Lambda \over R_\odot}\right)^{-1}
\label{eff}
\end{equation}
denotes the efficiency in extracting energy from the relativistic
outflow (mainly composed of $e^{\pm}$), and $\Psi(\varphi)=(3[\pi
-\varphi]- {1 \over 2}\sin 2\varphi - 4\sin \varphi )/(2\sin\varphi)$
is the beaming function of the generated $\gamma$-ray
radiation\footnote{Owing to the fact that the optical star is not a
point source, $\Psi(\varphi)$ is clearly only accurately for
$R_\ast/\delta < \varphi < \pi -R_\ast/\delta$, where $R_\ast$ is the
radius of the stellar companion.}. The total luminosity emitted by the
relativistic outflowing wind through the Compton-drag process in the
direction of the observer is highly anisotropic: $L_\Gamma (\varphi) =
[\Delta\gamma (\varphi)/ \gamma] L_\Omega$. $L_\Gamma$ strongly varies
with $\varphi$ which in turn changes periodically during orbital
motion (i.e. $\cos \varphi=\sin i \cos \nu$ for a circular orbit,
where $i$ is the inclination angle and $\nu$ is the true
anomaly). Fig. \ref{fig1} shows the luminosity carried by the
scattered photons for various positions of the companion and different
assumptions regarding the magnetar luminosity. 

\begin{figure*}
\centerline{\psfig{figure=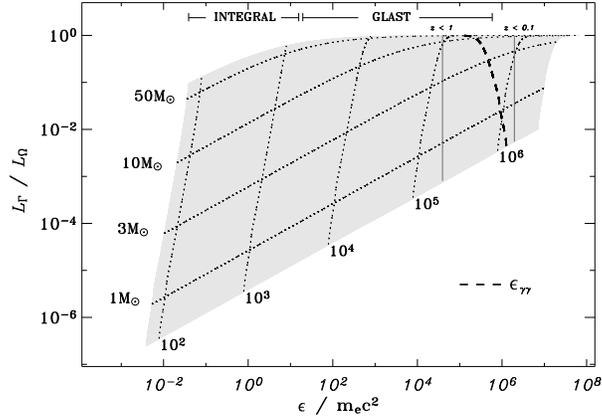,angle=0,width=0.55\textwidth}}
\vspace{-0.5cm}
{\caption{Efficiency in extracting energy from the relativistic
outflow as a function of the energy of the scattered $\gamma$-ray
radiation. The properties of the observed signal are linked to both
the evolutionary history of the companion and the bulk Lorentz factor
of the relativistic wind (i.e. $\gamma$). The main sequence radii and
luminosities of the companion star are calculated using the analytic
function derived by Tout et al. (1996). We assume $\Lambda=3R_*$. The
{\it dashed} line denotes the threshold energy at which the beamed
photons are absorbed by the soft field radiation. The two vertical
{\it solid} lines mark the lower limits to the maximum redshift to
which TeV $\gamma$-ray sources are visible, based on limits to the
background density of IR/UV photons (Biller et al. 1998). This has
been done for the $\gamma$-ray energy range of 0.4-10 TeV, assuming
that a source remains {\it visible} out to an optical depth of 2 and
that the majority of the IR was produced prior to the epoch of the
sources under study.}
\label{fig2}}
\end{figure*}

The Compton drag process can be very efficient in extracting energy
from the outflowing relativistic wind provided that $\varsigma \ge 1$
(see equation \ref{eff}). $\varsigma$ is of course uncertain, but this
number does not seem unreasonably high for a progenitor star with a
massive binary companion, and suggests that our fiducial value of
$L_\Omega$ for the overall luminosity need not be an overestimate
(Fig. \ref{fig2}). PSR B1259 -- 63 is an example of a system where the
mildly relativistic outflow from the aged pulsar is thought to
interact with the soft photon field radiation produced by its high
mass binary companion (Chernyakova \& Illarionov 1999). Since
$\varsigma$ could be less than $10^{-3}$ for a stellar companion with
modest luminosity (or a binary with $\Lambda \gg R_\ast$), this
suggests that we cannot rule out the possibility that part of the
$\gamma$-ray continuum could still come from the intrinsic pulsed
emission from the magnetar itself. For a typical aged pulsar this
mechanism yields $L_\Gamma / L_\Omega \sim 10^{-2}-10^{-3}$ (Arons
1996). The time dependence and spectral properties, in this latter
case, will be very distinctive and such effects should certainly be
looked for.

\subsection{Spectral attributes}

If the scenario proposed here is indeed relevant to an understanding
of the nature of GRB progenitors, then its existence becomes
inextricably linked to both the evolutionary history of the companion
and the characteristics of the hydromagnetic wind. The postburst
outflow from the compact remnant (which we assume to be highly
relativistic) propagates in the soft photon field of the stellar
companion where $U(r,\varpi)=m_ec^2 \varpi n_{\varpi}(r,\varpi)$ is
the photon energy density at the location $r$. As the stellar
companion emits a black body spectrum, of effective temperature
$\theta_\ast$, the local photon energy density is given by
\begin{equation}
n(r,\varpi)={2\pi \over h c^3}\left({m_e c^2 \over h}\right)^2
\left({R_* \over \delta[r]} \right)^2 {\varpi^2 \over
\exp[\varpi/\theta_\ast]-1},
\label{den}
\end{equation}
where $\varpi$ is the soft photon energy in units of $m_ec^2$. The
scattered photons are boosted by the square of the Lorentz factor so
that the local spectrum has a black body shape enhanced by $\gamma^2$.

\begin{figure*}
\centerline{ \psfig{figure=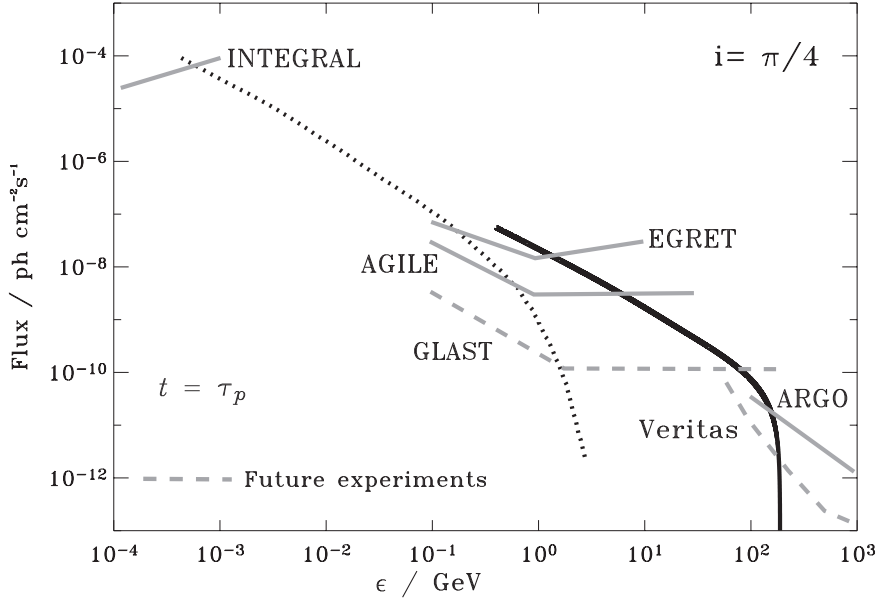,angle=0,width=0.68\textwidth}}
\vspace{-0.5cm} {\caption{Representative spectra produced by the
interaction of the relativistic outflowing wind with the radiation of
the stellar companion. The power output of the magnetically driven
relativistic wind declines as $(1+ t/t_{\rm m})^{-2}$, with $L_{{\rm
m},0} \simeq 10^{49} {\rm erg~s^{-1}}$, $t_{\rm m} \simeq 10^3~{\rm
s}$, $\zeta=1$, and $t=\tau_p$ = 12 hours (circular orbit). We assume
$z=0.5$ ($H_0 = 65\,\, {\rm km} \, {\rm s}^{-1} \, {\rm Mpc}^{-1}$,
$\Omega_{\rm m}=0.3$, and $\Omega_{\Lambda}=0.7$).  The luminosity of
the $\gamma$-ray emission is shown for $\gamma=10^4$ ({\it dotted
line}), $10^6$ ({\it solid line}), $\theta_\ast=2.5 \times 10^{-6}$,
and $\varsigma=1$. The thick grey lines are the predicted sensitivity
of a number of operational and proposed instruments in $\gamma$-ray
astrophysics (Morselli 2002).}
\label{fig3}}
\end{figure*}
The luminosity of the scattered emission moving along a radial
trajectory at an angle $\varphi$ is $L_\Gamma
(\gamma,\varphi,\epsilon)=\int_0^{\infty} \epsilon n_\gamma n_\varpi
(1- \beta \cos \omega) \sigma_{\rm KN} dV$, where $\epsilon$ is the
energy of the scattered photons in units of $m_ec^2$, and $\sigma_{\rm
KN}$ is the Klein-Nishina cross-section (Jauch \& Rohrlich 1976). As
can be seen in Fig. \ref{fig3}, the resulting spectrum is the
convolution of all the locally emitted spectra
(i.e. $\int_0^{\infty}\;dn_\varpi[r,\varpi]$) and it is not one of a
blackbody. Note that Klein-Nishina effects are important for incoming
photon energies such that $\varpi \gamma (1 + \cos \varphi)>1$.  The
maximum energy of the scattered photons in this regime is $\gamma
m_ec^2$.

A further effect which may strongly affect the observed spectrum is the
production of $e^{\pm}$ pairs through photon-photon collisions.
$e^{\pm}$ pairs can be produced by scattered photons interacting with
the isotropic companion emission or with each other. Photon collisions
within the beam itself occur between photons of equal age and can only
affect the high energy tail of the spectrum provided that $\gamma
\theta_* >1/3$ (Svensson 1987). Let us thus consider in turn the role
of scattered and companion radiation as seed photons for this
process. The interaction between the $\gamma$-rays produced by the
Compton-drag process and photons emitted by the companion star would
occur at large angles, resulting in an average energy threshold of
$\epsilon_{\gamma\gamma}>1/\gamma$. The radiation flux produced at the
location $r_\tau$ will then decreased by
$\exp[-\tau_{\gamma\gamma}(r_\tau,\epsilon)]$, where
$\tau_{\gamma\gamma} (r_\tau, \epsilon) =
\int_{\epsilon_{\gamma\gamma}}^{\infty}
d\varpi\int_{r_{\tau}}^{\infty}\sigma_{\gamma\gamma} (\varpi,\epsilon)
n_{\varpi} (r,\varpi) dr$ is the photon-photon optical depth and
$\sigma_{\gamma\gamma}(\varpi,\epsilon)$ is the corresponding
cross-section. As $\sigma_{\gamma\gamma}(\varpi,\epsilon)$ is peaked
at the threshold energy (Svensson 1987), the above expression can be
simplified to $\tau_{\gamma\gamma} (r_\tau, \epsilon) = (\sigma_T/
5)\int_{r_{\tau}}^{\infty} \epsilon_{\gamma\gamma}n_{\varpi}
(r,\epsilon_{\gamma\gamma})dr$, where
$n_\varpi(r,\epsilon_{\gamma\gamma})$ is the photon density at
threshold at the location $r$ (see equation \ref{den}).

As $\varpi \ll \gamma$, this absorption mechanism would be important
as long as the companion star produces a sufficient number of photons
with energies $\epsilon_{\gamma\gamma}>1/\gamma$. This limit is
illustrated in Fig. \ref{fig2} for various evolutionary histories of
the stellar companion, which has been assumed to be in the main
sequence (Izzard et al. 2003). It can be seen that the number of soft
photons able to interact with the high-energy $\gamma$-rays to produce
$e^{\pm}$ pairs strongly increases as the bulk Lorentz factor of the
relativistic wind exceeds $10^5$. The absorbed radiation will
subsequently be reprocessed by the pairs and redistributed in
energy. Each electron and positron will have an energy $\gamma_\pm
\sim \epsilon/2$ at birth, and will cool as a consequence of the
Compton-drag process. The positrons will in turn annihilate in
collisions with electrons in the wind, producing a blueshifted
annihilation line at $\epsilon \sim \gamma$.

Direct measurements of the characteristics of this hard-energy
radiation are frustrated by the fact that in traversing intergalactic
distances, $\gamma$-rays may be absorbed by photon-photon pair
production on the background field radiation (Gould \& Schr\'eder
1967). Photons of energy near 1 TeV interacting with background
photons of $\sim 0.5$ eV have the highest cross section, although a
broad range of optical-infrared wavelengths can be important absorbers
because the cross section for pair production is rather broad in
energy and, in addition, spectral features in the extragalactic
background density can make certain wave bands more important than the
cross section alone would indicate (Biller et al. 1998). The current
generation of ground-based, $\gamma$-ray telescopes have a typical
lower energy threshold of $\sim 0.5$ TeV and are thus expected to be
able to see sources possessing redshifts up to $z=0.1$. The next
generation of instruments (e.g. {\it Glast}\footnote{\tt
http://www-glast.slac.stanford.edu/}) is expected to have an energy
threshold in the region 0.05-0.1 TeV, and will therefore be able to
see out to a redshift of at least $z=0.5$. Fig. \ref{fig2} and its
caption summarise the above limits along with the spectral attributes
of the observed $\gamma$-ray signal.

\section{A decaying magnetar in an asymmetric SNR}

The success of the previous model is inseparably linked with the
assumed residence of a companion star. Here we consider a related and
less restrictive scenario in which the ambient photons are produced by
the postexplosion expansion of the disrupted envelope. Photons emitted
from the supernova remnant (SNR) shell provide copious targets for the
relativistic outflow to interact and produce high energy
$\gamma$-rays. 

GRBs are thought to be produced when the evolved core of a massive
star collapses either to a fast-spinning neutron star or a newly
formed black hole. In the latter case, a GRB is likely to be triggered
if the remaining star has sufficient angular momentum to form a
centrifugally supported disk (e.g. MacFadyen \& Woosley 1999).  A
funnel along the rotation axis would have been blasted open during the
1-100 s duration of the original burst; it would subsequently enlarge
owing to the postexplosion expansion of the envelope of the progenitor
star (e.g. Woosley 1993). The ram pressure of the continuing MHD
outflow would further enlarge the funnel. Besides asymmetrically
ejecting the SNR envelope (e.g., no ejecta in the polar direction),
the supernova explosion may leave behind parts of the He core which
take longer in falling back. Additional target photons may arise from
parts of the disrupted He core, no longer in hydrostatic equilibrium
and moving outwards inside the SNR shell. For a nominal
subrelativistic shell speed $v=10^9 v_9$ cm s$^{-1}$, the typical
distance reached is $r_{\rm snr} \approx 10^{14}v_9 t_{\rm d}$ cm in
$t_{\rm d}$ days. The outflowing wind from the compact remnant (which
we assume to be relativistic) would propagate inside the funnel cavity
(of conical shape with semi-aperture angle $\psi$) while interacting
with the SNR target photons with typical energy $\theta_{\rm snr}
\approx 10^{-6}-10^{-5}$ (i.e. optical frequencies).
\begin{figure*}
\centerline{\psfig{figure=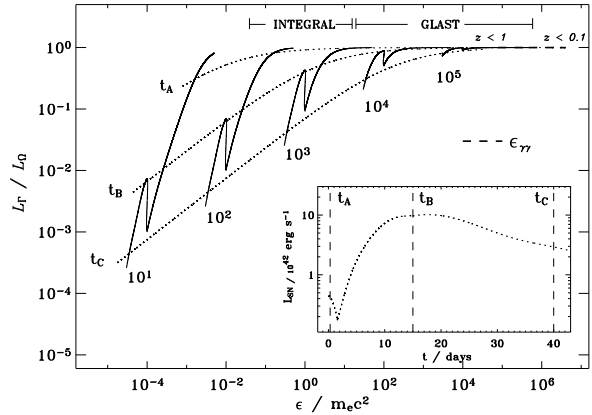,angle=0,width=0.55\textwidth}}
\vspace{-0.5cm} {\caption{Luminosity of the scattered $\gamma$-ray
radiation produced by the interaction of the postburst relativistic
outflow with the supernova photon field. The properties of the
observed signal are linked to the postexplosion expansion history of
the SN envelope. The bolometric luminosity of SN1998bw, as derived by
Woosley et al. (1999), is used to calculated the photon field energy
and column density (inset panel). We assume $v=10^9$ cm s$^{-1}$ and
$\psi=0.7$. The thick {\it dashed} line denotes the threshold energy
at which the beamed photons are absorbed by the radiation from the
SNR.}
\label{fig4}}
\end{figure*}

Under the foregoing conditions the particles in the postburst
relativistic outflow see blue-shifted photons pouring in from the
forward direction. The rate of energy loss of a relativistic particle
moving in a radiation with an energy density $w_{\rm snr}\approx
L_{\rm snr}/ 2\pi\psi^2r_{\rm snr}^2c$ is about $m_e c^2 d \gamma/dt
\approx - w_{\rm snr}\sigma_T c \gamma^2$ in the Thompson limit. The
total luminosity of scattered hard photons $L_\Gamma$ is equal to the
total particle energy losses in the course of motion from the magnetar
to infinity: $L_\Gamma (\theta_{\rm snr})=[\Delta \gamma(\theta_{\rm
snr})/\gamma]L_\Omega$. $L_\Gamma$ varies with $\theta_{\rm snr}$
which in turn evolves with the expansion history of the stellar
envelope. Fig. \ref{fig4} shows the total luminosity of scattered hard
photons $L_\Gamma$ as a function of the age of the remnant. The
bolometric luminosity of SN 1998bw, as derived by Woosley et
al. (1999), is used here to calculated the photon field energy by
assuming a constant expansion velocity of $v=10^9$ cm s$^{-1}$.  The
resulting radiation pressure on electrons in the ejecta will brake any
outflow whose initial Lorentz factor exceeds some critical value
$\gamma_{\rm cd} \leq (L_\Omega/L_{\rm snr})^{1/2}$, converting the
excess kinetic energy into a directed beamed of scattered photons. In
reality, the external parts of the postburst relativistic outflow,
which are in closer contact with the funnel walls, are dragged more
efficiently since the soft photons arising from the walls can
penetrate only a small fraction of the funnel before being upscattered
by relativistic electrons. The outflow itself, is then likely to
develop a velocity profile with higher Lorentz factor along the
symmetry axis, gradually decreasing as the polar angle
increases. Moreover, the outflow power may fluctuate; so also may its
baryon content, due to entrainment, or to unsteadiness in the
acceleration process at the base of the outflow. In this case,
additional $\gamma$-rays can be produced in relativistic shocks that
developed when fast material overtakes slower material (i.e. internal
shocks).

\section{Discussion}
The initial, energetic portion of the relativistic jet, with a typical
burst duration of 10 s, will rapidly expand beyond the envelope of the
progenitor star, leading in the usual fashion to shocks and a
decelerating blast wave. A continually decreasing fraction of energy
may continue being emitted for periods of days or longer, and its
reprocessing by the soft photon field radiation can yield a continuum
luminosity extending into the $\gamma$-ray band. Photons from a binary
companion and/or photons entrapped from a supernova explosion provide
ample targets for the wind to interact and produce high energy
$\gamma$-rays. With a sensitivity of $5 \times 10^{-5}$ ph cm$^{-2}$
s$^{-1}$ ($6 \times 10^{-9}$ ph cm$^{-2}$ s$^{-1}$) in a $\sim 10^5$ s
exposure, the $\gamma$-ray instrument on board of {\it
Integral\footnote{\tt http://sci.esa.int/home/integral/}} ({\it
Glast}) will detect a $10^{47}$ erg s$^{-1}$ signal out to a distance
of $z \sim 0.5$ ($\sim 1$). A magnetar with $P_0=0.8$ ms and $B\sim
10^{14}$ G would lead to $L_\Omega \sim 10^{47}$ erg s$^{-1}$ after 1
day. This could also be a consequence of a drop in $B$ from $10^{15}$
to $3 \times 10^{12}$ G in a compact structure with stored energy of
at least $10^{52}$ erg whose characteristic spin period remained
constant (at a fraction of a millisecond).  A similar argument can be
developed based on the concept of $\alpha$ viscosity. For a hot dense
torus around a black hole resulting from collapse of the core of the
progenitor star, the viscous accretion time for a torus of radius
$10^{9}r_9$ cm is $t_{\rm vis} \sim 1.5r_9^{3/2}B_{12.5}^{-2}$ for $nT
\sim$ constant, and the accretion of $\le 10^{-2}M_\odot$ in $t \sim
1$ day is sufficient to provide a characteristic $L \sim 10^{47}$ erg
s$^{-1}$. This luminosity may not dominate the continuum GeV afterglow
if it is only modestly reprocessed (i.e $L_\Gamma \le 10^{44}$ erg
s$^{-1}$). In this case, we cannot rule out the possibility that much
of the GeV emission is due to inverse Compton (M\'esz\'aros \& Rees
1994) or synchrotron self-Compton (Derishev et al. 2001) losses in the
standard decelerating blast wave, which could produce an afterglow
luminosity that was still, 1 day after the original explosion, as high
as $10^{45}$ erg s$^{-1}$ (see e.g. Dermer et al. 2000 for a blast
wave expanding into a medium with $n_0=100$ cm$^{-3}$).

In closing, the scattered $\gamma$-rays from a peculiar Ib/c SNR of
greater than usual brightness or a high luminosity stellar companion
(i.e. $> 10 M_\odot$) should be easily detectable (at $t_{\rm obs} \le
1.5$ days) by the recently launched {\it Integral}. For a less
luminous accompanying star (i.e. $< 5 M_\odot$), the expected
$\gamma$-ray flux can be detected up to $z=0.1$, although may be
difficult to disentangle from the scattered SNR emission whose
intensity is likely to dominate at $t \le 100$ days. The $\gamma$-ray
signals for ``mean'' events (i.e. $z \sim 1$) would stand out with
high statistical significance above the background provided that
$t_{\rm obs} < 0.1 - 0.5$ days. The planned {\it Glast} experiment may
also soon provide relevant limits (i.e. $z\le 0.5$) for individual
events at higher threshold energies (Figs. 2 and 3). Its detection
would unveiled the presence of a postburst relativistic wind, and the
spectral signatures of the $\gamma$-ray emission would help constrain
the nature of the precursor star.
 
\section*{Acknowledgements}
I gratefully acknowledge very helpful discussions with Martin J.~Rees
and Aldo Serenelli. I also thank the referee for valuable comments.
This research has been supported by NASA through a Chandra
Postdoctoral Fellowship award PF3-40028.

\end{document}